\documentstyle[epsfig,prd,aps,floats,amsfonts]{revtex}
\title{Nonsingular vacuum cosmologies with a variable
        cosmological term}
\author{K.A. Bronnikov,}
\address{VNIIMS, 3-1 M. Ulyanovoy St., Moscow 117313, Russia;\\
        Institute of Gravitation and Cosmology, PFUR, 6
        Miklukho-Maklaya St., Moscow 117198, Russia}
\author{A. Dobosz and I.G. Dymnikova}
\address{Department of Mathematics and Computer Science, University
    of Warmia and Mazury, \.Zolnierska 14, 10-561 Olsztyn, Poland}

\newcommand{\Ref}[1]{Ref.\,\cite{#1}}
\newcommand{\sect}[1]{Sec.\,#1}

\def\cm{\hspace*{1cm}}

\def\eq{Eq.\,}
\def\eqs{Eqs.\,}
\def\beq{\begin{equation}}
\def\eeq{\end{equation}}
\def\bear{\begin{eqnarray}}

\def\ear{\end{eqnarray}}

\def\dst{\displaystyle}

\def\nn{\nonumber\\ {}}

\def\eql{&=&}       

\def\e{{\,\rm e}}
\def\d{\partial}

\def\sign{\mathop{\rm sign}\nolimits}

\def\const{{\rm const}}
\def\Half{{\dst\frac{1}{2}}}

\newcommand{\aver}[1]{\langle \, #1 \, \rangle \mathstrut}

\def\Jl#1#2{{\it #1\/} {\bf #2},\ }

\def\CQG#1 {\Jl{Class. Qu. Grav.}{#1}}
\def\DAN#1 {\Jl{Dokl. AN SSSR}{#1}}
\def\GC#1 {\Jl{Grav. \& Cosmol.}{#1}}
\def\GRG#1 {\Jl{Gen. Rel. Grav.}{#1}}
\def\JETF#1 {\Jl{Zh. Eksp. Teor. Fiz.}{#1}}
\def\JETP#1 {\Jl{Sov. Phys. JETP}{#1}}
\def\JMP#1 {\Jl{J. Math. Phys.}{#1}}
\def\NPB#1 {\Jl{Nucl. Phys.}{B\ #1}}
\def\PLA#1 {\Jl{Phys. Lett.}{#1A}}
\def\PLB#1 {\Jl{Phys. Lett.}{#1B}}
\def\PRD#1 {\Jl{Phys. Rev.}{D\ #1}}
\def\PRL#1 {\Jl{Phys. Rev. Lett.}{#1}}

\def\mn{_{\mu\nu}}
\def\MN{^{\mu\nu}}
\def\mN{_\mu^\nu}

\def\L{{\mathbb L}}
\def\M{{\mathbb M}}
\def\R{{\mathbb R}}
\def\S{{\mathbb S}}
\def\V{{\mathbb V}}

\def\GR{general relativity}
\def\sph{spherically symmetric}
\def\ssph{static, spherically symmetric}
\def\fig{Fig.\,}
\def\bh{black hole}
\def\Sch{Schwarzschild}
\def\Lem{Lema\^\i tre}
\def\vac{{\rm vac}}

\def\dens{\mbox{g\,$\cdot$\,cm${}^{-3}$}}

\begin{document}

\twocolumn[\hsize\textwidth\columnwidth\hsize\csname
@twocolumnfalse\endcsname

\maketitle

\begin{abstract}
   We present nonsingular cosmological models with a variable
   cosmological term described by the second-rank symmetric tensor
   $\Lambda\mN$ evolving from $\Lambda \delta\mN$ to $\lambda
   \delta\mN$ with $\lambda<\Lambda$. All $\Lambda\mN$ dominated
   cosmologies belong to \Lem\ type models for an anisotropic perfect
   fluid. The expansion starts from a nonsingular nonsimultaneous de
   Sitter bang, with $\Lambda$ on the scale responsible for the
   earliest accelerated expansion, which is followed by an
   anisotropic Kasner type stage. For a certain class of observers
   these models can be also identified as Kantowski-Sachs models with
   regular R regions. For Kantowski-Sachs observers the
   cosmological evolution starts from horizons with a highly
   anisotropic ``null bang'' where the volume of the spatial section
   vanishes. We study in detail the spherically symmetric case and
   consider the general features of cosmologies with planar and
   pseudospherical symmetries. Nonsingular $\Lambda\mN$ dominated
   cosmologies are Bianchi type I in the planar case and hyperbolic
   analogs of the Kantowski-Sachs models in the pseudospherical case.
   At late times all models approach a de Sitter asymptotic with
   small $\lambda$.
\end{abstract}

\centerline{ PACS numbers: 04.70.Bw, 04.20.Gz, 98.80.Hw}
\vspace{6mm}

]  

\section{Introduction}

    Anything which contributes to the energy density of vacuum
    $\rho_{\vac}$ and satisfies the equation of state $p=-\rho$ (we
    adopt $c=1$ for simplicity), is associated with the Einstein
    cosmological term by \cite{gliner,zeld} (for
    reviews see \cite{adler,wein,cooper})
\beq
      T\mN=\rho_{\vac}\delta\mN=(8\pi G)^{-1}\Lambda\delta\mN   \label{vac}
\eeq
    The cosmological constant $\Lambda$ must be constant by the conservation
    equation $\nabla_\nu T\mN =0$ that follows from the contracted Bianchi
    identities $\nabla_\nu G\mN=0$ and the Einstein equations $G\mN =
    - 8\pi G T\mN$.

    Developments in particle and quantum field theories as well as
    inflationary scenarios and observational cosmology compellingly
    suggest that the cosmological constant $\Lambda$ has to be a dynamical
    quantity.

    In inflationary models in which an accelerated expansion due to
    $\Lambda$ is based on generic properties of the de Sitter vacuum
    (\ref{vac}), independently of where $\Lambda$ comes from \cite{gd75},
    various mechanisms have been proposed relating the cosmological term
    $\Lambda \delta\mN$ to matter sources (for reviews see
    \cite{linde,olive}).  A huge cosmological
    constant or vacuum density at the earliest stage of the Universe
    evolution ($\rho_{\vac} \sim 10^{77}$ \dens\ for the Grand
    Unification scale $E_{\rm GUT}\sim{10^{15}}$ GeV) provides an accelerated
    expansion needed to guarantee the survival of the Universe to its
    present size and density \cite{gd75} as well as to explain its observed
    homogeneity and isotropy and solve other puzzles of the standard Big
    Bang cosmology (for a review and a list of puzzles see
    \cite{linde,olive}). On the other hand, cosmological observations
    require the present value of the vacuum density to provide a
    substantial contribution (about 70 per cent) to the total density
    in the Universe, $\rho\sim 10^{-30}$ \dens\ \cite{bahcall}.

    Most of the models for $\Lambda$ dynamics considered in the literature
    are of ``cancelling type'' (for reviews see \cite{wein,cooper,turner}),
    involving fields (most frequently scalar fields non-minimally coupled to
    gravity \cite{dolgov,ruggi}) which develop a negative energy density
    growing with time to ultimately cancel a pre-existing (driving
    inflation) positive value of $\Lambda$. This approach
    typically requires the gravitational constant $G$ to be time-
    \cite{dolgov} or scale- \cite{bonnano} dependent (the question of
    confronting $G$-variable models with observations is addressed in
    Ref.\,\cite{alosha}). Models with a dynamical $\Lambda$-term
    depending binomially on the scalar field are proposed in the context
    of scalar-tensor theories with matter described as a perfect fluid
    with a barotropic equation of state \cite{diaz}.

    Another wide class of models are phenomenological FRW (homogeneous and
    isotropic) cosmologies with an effective stress-energy
    tensor (SET) describing both the $\Lambda$-term and non-vacuum matter as
    perfect fluids (for a review see \cite{cooper}).
    In Ref.\,\cite{cooper}, a number of new exact singular
    FRW solutions are presented for the case $\Lambda \propto t^{-l}$ and
    $k=0$ (spatially flat isotropic models), with $\Lambda <0$ for odd
    values of $l$; several numerical models are built for $k=1$ (a closed
    universe) with $\Lambda$ varying as a function of the scale factor $a$,
    and for the case of an open universe ($k=-1$) with $\Lambda$ varying as
    a function of the Hubble parameter $H$. The paper \cite{lima} presents a
    nonsingular isotropic cosmology with a phenomenological ansatz
    postulating $\Lambda$ decay from an initial de Sitter state which
    ideologically resembles the old nonsingular FRW model with the de Sitter
    vacuum as an initial state \cite{gd75}.

    In Ref.\,\cite{vish}, plane-symmetric Bianchi type I models are
    considered with a perfect fluid source including dust, radiation and
    time-dependent $\Lambda$. The evolution starts from an initial
    singularity followed by a highly anisotropic stage which involves a big
    negative value of $\Lambda$ at the beginning in the case of the critical
    density \cite{vish}.  In Ref.\,\cite{chim}, isotropic but inhomogeneous
    spherically symmetric cosmologies are considered with a material fluid,
    a heat flux, a $Q$ matter (related to a self-interacting scalar field)
    and a time-dependent cosmological constant $\Lambda$; late-time
    izotropization depends on the time evolution of an effective adiabatic
    index for the material fluid \cite{chim}.

    In this paper we present nonsingular cosmological models with a
    variable cosmological term described by the second-rank symmetric tensor
    $\Lambda\mN$ \cite{id00} in which a constant $\Lambda$ associated with
    vacuum density by (\ref{vac}), becomes a tensor component
    $\Lambda_t^t=(8\pi G) T_t^t$ associated explicitly with the density
    component of the perfect fluid SET $T\mN=(8\pi G)^{-1}\Lambda\mN$
    \cite{id00} whose vacuum properties follow from its symmetry and whose
    variability follows from the Bianchi identities \cite{id92}.
    This approach allows us to specify possible types of cosmological models
    with variable vacuum density consistent with the Bianchi identities.

    Let us emphasize that our approach is phenomenological in essence.

    In quantum field theory, the symmetry of the vacuum expectation
    value of the SET $\aver{T\mn}$ does not always coincide with the
    symmetry of the background space-time, since quantum field theory in
    curved space-time does not contain a unique specification of the quantum
    state of a system \cite{bird,gmm,mottola}. Even in Minkowski space-time
    there are vacuum states (e.g., the well-known Rindler vacuum) which are
    non-invariant under the full Poincar\'e group. In the case of de Sitter
    space the renormalized expectation value of the vacuum SET $\aver{T\mn}$
    for a scalar field with an arbitrary mass $m$ and curvature coupling
    $\xi$ is proved to have a fixed point attractor behavior at late times
    \cite{mottola} approaching, depending on $m$ and $\xi$, either the
    Bunch-Davies de Sitter-invariant vacuum \cite{bunch} or, for the
    massless minimally coupled case ($m=\xi=0$), the de Sitter-invariant
    Allen-Folacci vacuum \cite{AF}. The case $m=\xi=0$ is peculiar (see,
    e.g., \cite{kirga}), since the de Sitter-invariant two-point function
    is infrared divergent, and the vacuum states free of this divergence are
    $O(4)$-invariant Fock vacua introduced by Allen \cite{allen}.  The
    vacuum energy density in the O(4)-invariant case is not the same
    (larger) than in the de Sitter-invariant case \cite{kirga}.

    The problems of specifying vacuum states in quantum field theory
    are beyond the scope of this paper.

    We here address the question: what can be the further cosmological
    evolution for cases when the vacuum density (no matter where it came
    from) had once contributed to a SET of the form (\ref{vac}).

    Phenomenologically, the SET (\ref{vac}) is defined as a vacuum SET due
    to its maximally symmetric form invariant under any coordinate
    transformations which makes impossible to single out a preferred
    comoving reference frame \cite{gliner,press}. As a result, an observer
    moving through a medium with a SET of structure (\ref{vac}) cannot in
    principle measure his velocity with respect to it, which allows one
    to classify it as a vacuum \cite{gliner,ll}.

    It is clear that if one wants to make variable a vacuum density defined
    in this way, one cannot retain the full invariance of the vacuum SET.
    The invariance can, however, be partial, valid in a certain Lorentzian
    subspace of the space-time manifold. In this case, the full symmetry is
    reduced but the invariance is still present for an observer moving along
    a certain direction in space.

    This approach has been proposed in Ref.\,\cite{id92} (for a review see
    \cite{id97,id99}), where a spherically symmetric vacuum has been
    introduced as defined by the algebraic structure of its SET such that
\beq
       T_t^t=T_r^r, \cm
       T_{\theta}^{\theta}=T_{\phi}^{\phi}              \label{Tstru}
\eeq
    We adopt the metric signature $(+---)$. The energy density is given by
    $\rho=T^t_t$, the radial pressure is $p_r=-T_r^r$ and the transversal
    pressure is $p_{\perp}=-T_{\theta}^{\theta}=-T_{\phi}^{\phi}$.

    The vacuum properties of (\ref{Tstru}) follow from its  invariance under
    boosts in the radial direction which results in the absence of a
    preferred comoving reference frame and makes impossible for an
    observer to measure the radial component of his velocity \cite{id92}.

    The vacuum tensor of this kind was first introduced as a source term for
    an exact spherically symmetric solution describing a black hole whose
    singularity is replaced with a de Sitter vacuum core \cite{id92,idea}.
    Later it was shown that the existence of the class of solutions to the
    Einstein equations with the source term of the form (\ref{Tstru}),
    asymptotically de Sitter at the center, is distinguished by the dominant
    energy condition and by the requirements of regularity of density and
    finiteness of the ADM mass \cite{id02}.  A SET from this class smoothly
    connects two de Sitter vacua --- at the center and at infinity, and
    corresponds to an extension of the Einstein cosmological term $\Lambda
    \delta\mN$ to the second-rank symmetric tensor with the algebraic
    structure (\ref{Tstru}) \cite{id00},
\beq
    \Lambda\mN = 8\pi G T\mN{}                           \label{lmunu}
\eeq
    evolving from $\Lambda \delta\mN$ as $r\to 0$ to $\lambda \delta\mN$ as
    $r\to \infty$, with $\lambda < \Lambda$. The Bianchi identities result
    in the conservation equation $\Lambda^{\mu}_{\nu;\mu}=0$ which gives the
    $r$-dependent equation of state \cite{id00}
\beq
    p_r^{\Lambda}=-\rho^{\Lambda}, \qquad
    p_{\bot}^{\Lambda} = -\rho^{\Lambda}-
                    \frac{r}{2}\frac{d\rho^{\Lambda}}{dr}     \label{eqst}
\eeq
    where $\rho^{\Lambda} (r) = (8\pi G)^{-1}\Lambda^t_t$,
    $p_r^\Lambda =-(8\pi G)^{-1} \Lambda_r^r$ and
    $p_{\perp}^{\Lambda}=-(8\pi G)^{-1}\Lambda_{\theta}^{\theta}=
    -(8\pi G)^{-1}\Lambda_{\phi}^{\phi}$. The global structure of the
    $\Lambda\mN$ geometry contains black and white holes whose singularities
    are replaced with regular cores, asymptotically de Sitter as $r\to 0$.
    A regular core of a white hole considered in the  cosmological
    coordinates ($R, \tau$), models the early stages of an expanding
    universe dominated by $\Lambda\mN$; the  evolution starts from a
    nonsingular non-simultaneous de Sitter bang (a bang is defined by
    $r(R,\tau)=0$ \cite{silk}) followed by a Kasner-type stage of
    anisotropic expansion at which most of the mass is produced \cite{us2001}.

    In  this paper we study $\Lambda\mN$-dominated cosmologies in a more
    general cosmological context. Namely, if there is a spatial direction
    distinguished by the symmetry of the source, we can choose a class of
    coordinate frames where this direction is parametrized by a certain
    coordinate $u$. Then our choice of the vacuum SET is only restricted by
    the requirement that it should be invariant under coordinate
    transformations in the $(u,t)$ subspace, or, in other words, under
    boosts in the distinguished two-dimensional Lorentzian subspace
    $\M^2(u,t)$ of the four-dimensional space-time manifold. It is then
    natural to assume that the two-dimensional subspaces orthogonal to
    $\M^2$ are maximally symmetric, i.e., spaces of constant curvature. The
    latter can be spheres $\S^2$, planes $\R^2$ or Lobachevsky planes
    $\L^2$, attached to each point of $\M^2$ with a certain scale factor
    which depends on a point in $\M^2$, i.e., in general on both coordinates
    $u$ and $t$.

    The paper is organized as follows. Sec. II presents the basic equations
    and exact solutions with a variable cosmological term (\ref{lmunu})
    for the cases of spherical, planar and pseudospherical symmetries.
    Sec. III is devoted to a systematic study  of \sph\ models.
    Conclusions are summarized in Sec. IV.

\section{$\Lambda_{\mu}^{\nu}$ dominated geometries}\label{s2}

\subsection{Basic relations}

    We consider 4-spaces with the structure
\beq
      \M^4 = \M^2 \times \V^2, \quad
      \V^2 = \left\{ \matrix{
            \S^2, &\ K=1;\cr
            \R^2, &\ K=0;\cr
            \L^2, &\quad K=-1.\cr}\right.               \label{Mstru}
\eeq
    Here $\M^2$ is the Lorentzian 2-subspace parametrized by the coordinates
    ($u,t$); $\V^2$ is a 2-surface with spherical ($K=1$), planar ($K=0$) or
    pseudospherical ($K=-1$) symmetries ($\phi$ and $\theta$ in
    (\ref{Tstru}) are any coordinates parametrizing $\V^2$). In the
    cosmological context, all three symmetries can exist on equal rights.
    (The index $K$ should not be confused with $k=0,\pm 1$, the spatial
    curvature index of the FRW models.)

    An arbitrary static metric for the source term (\ref{Tstru}) and
    the space-time structure (\ref{Mstru}) is given by
\beq
    ds^2 = A(u) dt^2 - \frac{du^2}{A(u)} - r^2(u)  d\Omega_K^2, \label{ds}
\eeq
    where $d\Omega_K^2$ is the squared linear element on a unit sphere
    ($K=1$), plane ($K=0$) or Lobachevsky plane ($K=-1$).

    The choice of the ``radial'' coordinate $u$ such that $g_{tt}g_{uu}=-1$
    is not only one of the widely used coordinate gauges: it is especially
    convenient when considering Killing horizons which correspond to zeros
    of the function $A(u)$. The reason is that in a close neighborhood of a
    horizon the coordinate $u$ varies (up to a positive constant factor)
    like manifestly well-behaved Kruskal-like coordinates used for an
    analytic extension of the metric (\ref{ds}). Therefore one can jointly
    describe in terms of $u$ the metric on both sides of a horizon. In
    addition, with the same coordinate, horizons correspond to regular
    points in geodesic equations \cite{cold}.

    It should be emphasized that the static form chosen for the metric
    (\ref{ds}) is not a restriction. As soon as we postulate the space-time
    structure (\ref{Mstru}) and the SET structure (\ref{Tstru}), the
    generalized Birkhoff theorem \cite{brkov,bm95} guarantees the
    existence of a coordinate frame where the metric has the form
    (\ref{ds}). The only restriction is the assumption that, in the
    space-time domain under consideration, the metric coefficient $r^2(u,t)$
    is not constant and has a non-null gradient:
    $\sign (g\MN r_{,\mu} r_{,\nu}) = \pm 1$. If it is $+1$, we are dealing
    with a T (cosmological) region, and $-1$ corresponds to an R
    (static) region.

    Indeed, according to (\ref{ds}), $g\MN r_{,\mu} r_{,\nu} = -A (dr/du)^2$.
    The space-time regions where $A > 0$ are static, the $u$ coordinate
    being spatial, and are called R regions. In T regions, where
    $A<0$, the coordinates $u$ and $t$ interchange their roles: $u$ becomes
    timelike and $t$ spacelike. In the \sph\ case, as is well known, the
    metric in a T region describes a Kantowski-Sachs (KS) type homogeneous
    anisotropic cosmological model \cite{kompa,kant}, representing a special
    case of T-models \cite{Ruban}, which are, in general,
    inhomogeneous. A spatial section of a KS model has the structure $\R
    \times \S^2$, a 3-dimensional cylinder with different time-dependent
    scale factors in the spherical and longitudinal directions.

    In the case of planar symmetry ($K=0$), in a T region we obtain
    anisotropic models with planar spatial sections, i.e., Bianchi type I
    models, more precisely, their plane-symmetric subset where two of the
    three scale factors coincide.

    In the pseudospherical case ($K=-1$), spatial sections in a T-region
    have the structure $\R \times \L^2$, similar to KS models but with
    spheres replaced by Lobachevsky planes. We will call such models
    HKS (hyperbolic Kantowski-Sachs) models.

    In regions where $A(u) < 0$, it is convenient to change the
    notation:  $t\mapsto x$ and $-A(u) \mapsto b^2(u)$, remembering that $u$
    is now a temporal coordinate. The metric is then rewritten as
\beq
    ds^2
      = \frac{1}{b^2(u)}du^2 - b^2(u)dx^2 - r^2(u) d\Omega_k^2, \label{dsT}
\eeq
    which describes an anisotropic cosmological model with two
    time-dependent scale factors $b(u)$ and $r(u)$ and the lapse function
    $1/b^2(u)$. A transition to synchronous time $\tau$ is performed using
    the integral
\beq
     \tau (u) = \int \frac{du}{b(u)}.                          \label{tau1}
\eeq

    A horizon $u=h$ of order $n$ is a zero of the same order of the
    function $A(u)$. For the
    metric (\ref{dsT}), the horizon $u=h$ is a coordinate
    singularity where the metric coefficient $g_{xx}$ vanishes, so that
    coordinate surfaces (e.g., spheres in case $K=1$) with the same finite
    scale factor $r(h)$ stick to one another.  By \eq (\ref{tau1}), this
    happens at finite cosmological time $\tau$ for a simple (first-order)
    horizon and in an infinitely remote past or future ($\tau \to \pm
    \infty$) for higher-order horizons.

    On the other hand, if a T region is located at large $r$, the anisotropic
    cosmology can isotropize at late times $\tau$ under the condition
    $b(\tau)\propto r(\tau)$. One can notice that for $K=\pm 1$ the
    isotropization can be only local. Indeed, for $K=1$ the spatial topology
    is cylindrical, and the directions along and across the coordinate
    spheres are non-equivalent.  For $K=-1$ the spatial topology is flat,
    but the global geometry is different along and across the Lobachevsky
    planes.

    Two independent components of the Einstein equations and the
    conservation equation for $\Lambda\mN$ give the dynamical equations
\bear
    2A r''/r = \Lambda^u_u - \Lambda^t_t \eql 0,              \label{e01}
\\
    \frac{1}{r^2} \bigl( -K + Ar'^2 + A'rr' \bigr) \eql
                       - \Lambda_t^t =  - 8\pi G \rho,        \label{e11}
\\
    r\rho' + 2(\rho + p_\bot) \eql 0,                         \label{cons}
\ear
    where the prime denotes $d/du$.

    \eq (\ref{e01}) leads to $r'' =0$. Solutions with $r = \const$ can be of
    certain interest, but not in the cosmological context since they cannot
    describe an expanding universe with isotropization at late times. Let us
    note that in case $r=\const$ the generalized Birkhoff theorem does not
    work and, in addition to solutions with the metric (\ref{ds}), there
    exist wave solutions \cite{brkov,bm95}. Here we do not consider such
    solutions.

    Now without loss of generality we can put $u \equiv r$ and consider the
    remaining unknowns as functions of $r$. We are left with two equations
    (\ref{e11}) and (\ref{cons}) for three unknowns $A(r)$, $\rho(r)$,
    $p_\bot(r)$. The set of equations becomes determined if we postulate an
    equation of state connecting $\rho$ and $p_\bot$ or specify the density
    profile $\rho(r)$.

    \eq (\ref{e11}) can be integrated giving
\beq
    A(r) = K -\frac{2G M(r)}{r},                     \quad
    M(r) = 4\pi \int_{d}^r \rho(x) x^2 dx.                 \label{AM}
\eeq
    with an arbitrary constant $d$. This is a solution in quadratures
    for any given $\rho(r)$, $p_\bot$ is then determined from (\ref{cons}).

    In the particular case of the equation of state $p_\bot = -\rho$, which
    leads to the usual cosmological constant, $\lambda = 8\pi G\rho =
    \const$, the integration (\ref{AM}) gives
\beq
    A(r) = K - \frac{2Gm}{r} - \frac{1}{3}\lambda r^2      \label{SdS}
\eeq
    which is the extension of \Sch-de Sitter geometry (described by the
    Kottler-Trefftz solution \cite{kot}) to the cases $K=0, -1$; $m$ is an
    integration constant ($m=0$ when $d=0$). It is interpreted as the mass
    in case $K=+1$, $\lambda=0$ (the \Sch\ solution).

\subsection {De Sitter space-time}

    This is the maximally symmetric solution to the Einstein equations with
    the cosmological constant $\Lambda >0$.
    It is the special case $m=0$ of the solution
    (\ref{ds}), (\ref{SdS}), so that the metric is
\beq
    ds^2 = A(r) dt^2 - \frac{dr^2}{A(r)} - r^2 d\Omega_k^2,  \label{ds_r}
\eeq
    $A(r)=K - H^2 r^2$, where $H^{-1} = \sqrt{3/\Lambda}$ is the curvature
    radius of this constant-curvature manifold. Different values of $K$
    correspond to the description of the same space-time in different
    coordinate frames. Let us briefly discuss the
    properties of the  metric (\ref{ds_r}) since the textbook descriptions
    (see, e.g., \cite{bird,hawkell}) discuss its \ssph\ form (which is the
    $K=1$ case in our notation) and its FRW forms, while we also need the de
    Sitter metric in KS and HKS forms (\ref{dsT}) to specify the asymptotic
    behavior of our models.

    In case $K=0$, a static region is absent ($A < 0$), and
    the metric (\ref{dsT}) transforms into the FRW form \cite{hawkell}
\beq
     ds^2 = d\tau^2 - \e^{2 H\tau} (dx^2 + dy^2 + dz^2),   \label{dS_0}
\eeq
    The expansion begins at $\tau \to -\infty$, i.e., in the infinitely
    remote past of cosmological observers.

    In the \sph\ case, in the nonstatic region $Hr > 1$  the KS metric
    for de Sitter space takes the form
\bear
    ds^2 \eql (H^2 r^2-1)^{-1} dr^2 - (H^2 r^2 -1)\,dx^2 - r^2 d\Omega_1^2
\nn
    \eql d\tau^2 - \sinh^2(H\tau)\,dx^2
                      - H^{-2} \cosh^2(H\tau)d\Omega_1^2.      \label{dS+}
\ear
    The expansion in this model begins at $\tau=0$ from a highly anisotropic
    apparently singular state with $g_{xx}=0$ and becomes locally isotropic
    (as noticed in Sect. IIA) and exponential (with the Hubble constant $H$)
    at large values of $\tau$.

    In case $K=-1$, the HKS metric (\ref{dsT}) in terms of $\tau$ reads
\beq
    ds^2 = d\tau^2 - \cosh^2(H\tau)\,dx^2
                      - H^{-2} \sinh^2(H\tau)d\Omega_{-1}^2.   \label{dS-}
\eeq
    The expansion begins with an anisotropic apparent singularity at
    $\tau=0$, where the Lobachevsky planes are drawn to points. The further
    expansion, as in other cases, ultimately becomes isotropic and
    exponential.

    Apparent singularities at the start of expansion are in all three cases
    produced by the choice of coordinate frames since the de Sitter geometry
    is globally regular and maximally symmetric. All 4-points of the de
    Sitter manifold (including those which seem to be singular in particular
    representations of the metric) are equivalent to each other.

    Let us note that, comparing the KS and HKS forms of the de Sitter metric
    with its FRW forms, we see that the topology of spatial sections
    of the same de Sitter 4-geometry depends on the choice of the coordinate
    frame.  Indeed, the spatial topology is $\R^3$ for an open and $\S^3$
    for a closed FRW model, $\R\times\S^2$ for the KS representation
    (\ref{dS+}) and  $\R^3$ for the HKS form.

    So far we discussed expanding models with the de Sitter metric. Their
    contracting counterparts are easily obtained by time reversal ($\tau \to
    -\tau$).

\begin{figure}
\vspace{-8.0mm}
\begin{center}
\epsfig{file=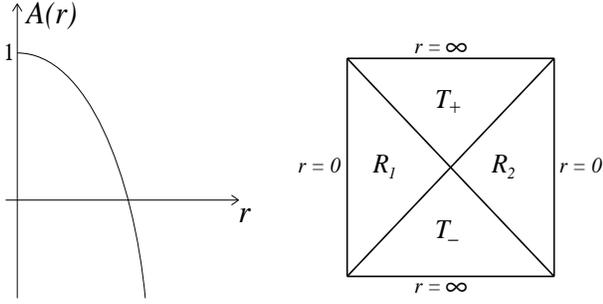,width=8.0cm,height=4.0cm}
\end{center}
\caption{De Sitter geometry.}
\label{fig.1}
\end{figure}

    \fig 1 shows a plot of $A(r)$ and the Carter-Penrose global structure
    diagram for de Sitter space-time. The metrics (\ref{dS+}), (\ref{dS-})
    describe either the contracting region $T_-$ for a contracting Universe
    ($\tau\in \R_-$), or $T_+$, for an expanding Universe ($\tau\in \R_+$).
    (R regions admit both expansion and contraction.)

\subsection{Cosmologies with a variable cosmological term}

    Consider now the general case of the solutions (\ref{ds}), (\ref{AM})
    with the cosmological term $\Lambda\mN$ of the algebraic structure
    (\ref{Tstru}). The solutions of the class to be considered satisfy the
    dominant energy condition ($\rho \geq 0,\ |p_i| \leq \rho,\ i=1,2,3$),
    are regular at $r\to 0$ and tend to the de Sitter metric
    with a finite value of the density $\rho$ as $r\to \infty$ \cite{id02}.
    The derivative of the density is $\rho' = -2(\rho + p_\bot)/r$ by
    (\ref{cons}),  and $\rho' < 0$ follows from the dominant energy
    condition, so that $\rho(r) = 8\pi G\Lambda ^t_t$ is a monotonically
    decreasing function of $r$.  This guarantees the existence of
    cosmological solutions evolving from a large value of the cosmological
    constant at small $r$ to its small value at large $r$.

    It can be verified by using the Kretchmann scalar
    ${\cal R}^2=R_{iklm}R^{iklm}$, where $R_{iklm}$ is the Riemann curvature
    tensor, that, for the class of metrics under consideration, regularity
    at $r=0$ is only achieved if $A(r)$ behaves at small $r$ as $k - A_2 r^2
    + o(r^2)$, where $A_2$ is a non-negative constant. This means that any
    solution with $\Lambda\mN$, regular at $r=0$, is asymptotically de
    Sitter at small $r$ (with $A_2 = H^2$ in the above notations) not only
    in the spherically symmetric case \cite{id02}, but also in the cases of
    planar and pseudospherical symmetries.

    According to (\ref{ds_r}), the spherical model ($K=1$) contains a regular
    R region near $r=0$, while the planar and pseudospherical models ($K=0,
    -1$) contain regular T regions near $r=0$.

    On the other hand, \eq (\ref{e11}) implies that on a horizon,
    where $A =0$,
\beq
    r A' = K - 8\pi G \rho r^2.                          \label{A'hor-}
\eeq
    For $K=0, -1$ this means that $A' <0$ at any horizon, i.e., it must be
    a simple horizon leading, as $r$ increases, from R to T region.  Since,
    as we have seen, regularity at $r=0$ requires that the region near $r=0$
    is a T region (with a de Sitter behavior), we
    have to conclude that all models considered here with $K=0$
    and $K= -1$ have no horizon, comprise T regions only and are
    purely cosmological. They describe anisotropic evolution between states
    with large and small values of cosmological constant, but effects
    connected with the existence of horizons are absent.

    In case $K=0$ the model is classified as a Bianchi type I model
    and in case $K=-1$ as a hyperbolic Kantowski-Sachs model.

    Spherically symmetric systems are much more diverse and complicated,
    and we discuss them in more detail in the next section.

\section {Spherically symmetric models}  \label{s4}

\subsection{One- and two-lambda configurations}

    In the \sph\ case, $K=1$, the metric has the form
\beq
    ds^2 = A(r) dt^2 - \frac{dr^2}{A(r)}
           - r^2 (d\theta^2 + \sin^2\theta\,d\phi^2).      \label{g_sph}
\eeq

    Requiring regularity at the center $r=0$, we put $d=0$ in (\ref{AM}),
    and $M(r)$ is, as usual, interpreted as the mass inside a sphere of
    radius $r$. Near the center the solution approaches the de Sitter metric
    with $\Lambda= 8\pi G\rho(0)$ \cite{id02}.  The function $A(r)$ is given
    by
\beq
    A(r) = 1 -\frac{2G  M(r)}{r},
    \quad M(r) = 4\pi \int_0^r \rho(x) x^2 dx.               \label{A,M}
\eeq

    If we require asymptotic flatness, then the monotonically decreasing
    function $\rho(r)$ should vanish as $r\to \infty$ quicker than $r^{-3}$,
    and the total gravitating mass (ADM mass) $m = M(\infty)$ is finite
    \cite{id02}. The de Sitter-Schwarzschild geometry, asymptotically de
    Sitter as $r\to 0$ and asymptotically Schwarzschild as $r\to \infty$,
    describes a vacuum nonsigular black hole for masses $m\geq m_{\rm crit}$
    where $m_{\rm crit}$ is a critical value, and a particlelike structure
    without horizons for $m < m_{\rm crit}$ in the Minkowski background
    \cite{id92,id96} (see \fig 2).

\begin{figure}
\vspace{-8.0mm}
\begin{center}
\epsfig{file=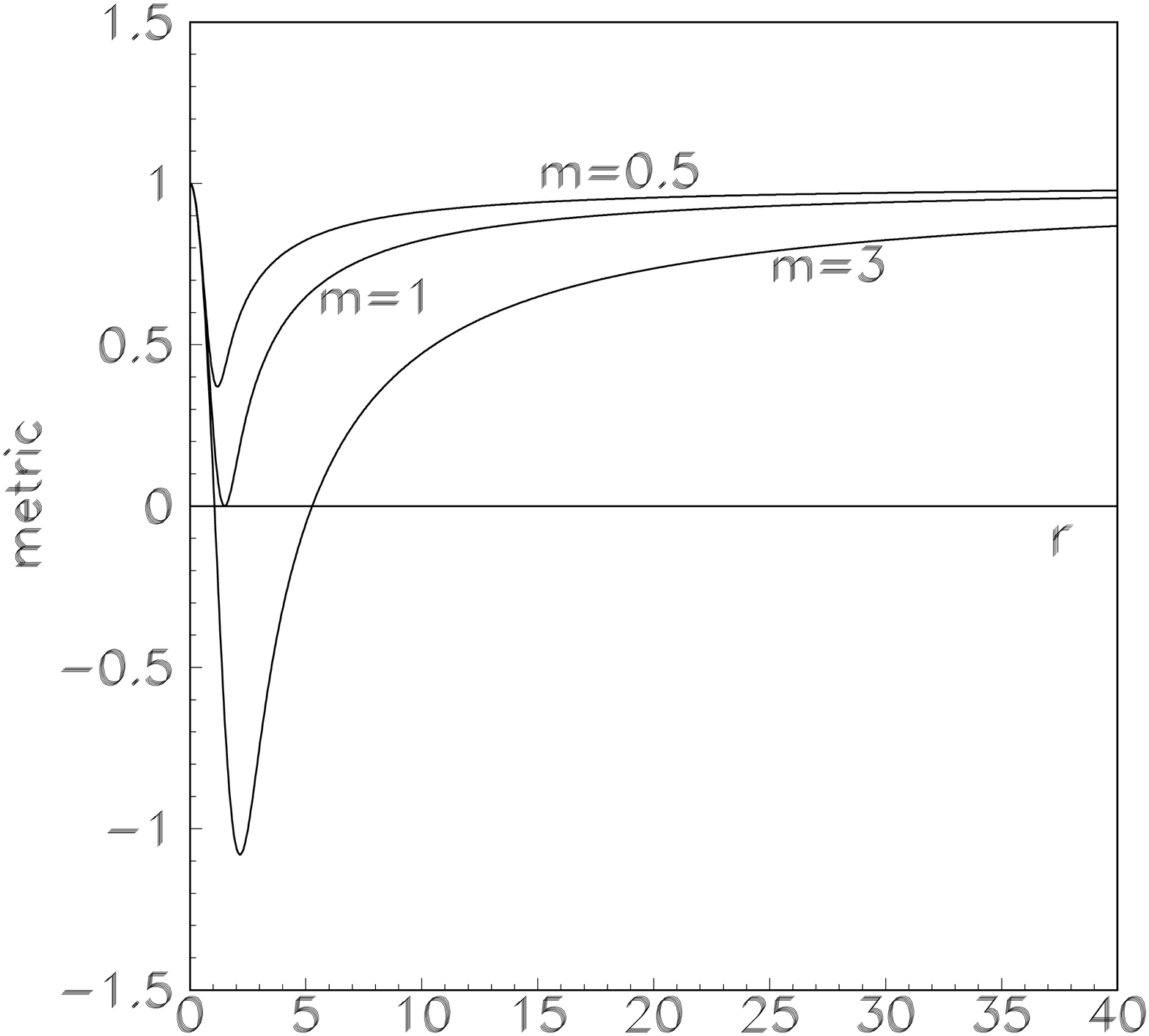,width=8.0cm,height=5.5cm}
\end{center}
\caption{The metric function $A(r)$ for the de
Sitter-Schwarzschild geometry. The mass parameter is normalized to
a critical mass.}
\label{fig.2}
\end{figure}

    The global structure of de Sitter-Schwarzschild space-time for $m\geq
    m_{\rm crit}$  contains an infinite set of vacuum nonsingular black and
    white holes whose singularities are replaced by regular cores with de
    Sitter vacuum near $r = 0$ (see \fig 3).
\begin{figure}
\vspace{-3.0mm}
\begin{center}
\epsfig{file=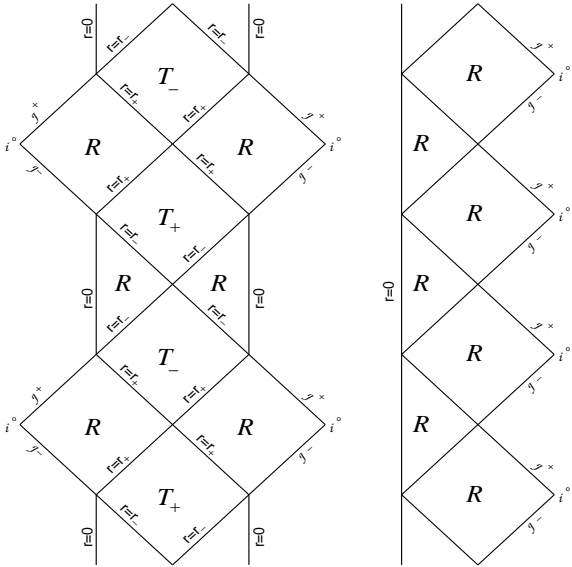,width=8.0cm,height=8cm}
\end{center}
\caption{Global structure of de Sitter-Schwarzschild space-time with two
    horizons (left) and one double horizon (right).}     \label{fig.3}
\end{figure}

    The regular core $R$ of a white hole $T_{+}$ models a nonsingular start
    of cosmological evolution which we consider in the next section.

    This geometry is easily extended to the case of a nonzero cosmological
    constant $\lambda < \Lambda$ at infinity \cite{us97}. The solution is
    described by the metric function \cite{us97}
\beq
      A(R)=1-\frac{2GM(r)}{r}-\frac{\lambda}{3}r^2 \label{sd}
\eeq
    with $\rho \to \rho_{\infty} = \lambda/(8\pi G)$ as $r\to {\infty}$. For
    a particular density profile of this kind \cite{id92},
\beq
    \rho (r) = \rho_0 \exp (-r^3/r_1^3) + \rho_\infty     \label{rho}
\eeq
    (where $\rho_0,\ r_1$ and $\rho_\infty$ are constants), $A(r)$ is
    shown in \fig 4, where $q=\sqrt{\Lambda/\lambda}$ \cite{us97}. The
    cosmological term $\Lambda\mN$  smoothly connects two de Sitter vacua
    with different values of the cosmological constant.  The geometry
    describes five types of globally regular configurations with
    qualitatively different behavior of $A(r)$ (see \fig 4
    \cite{us97,us98}).

\begin{figure}
\vspace{-2.0mm}
\begin{center}
\epsfig{file=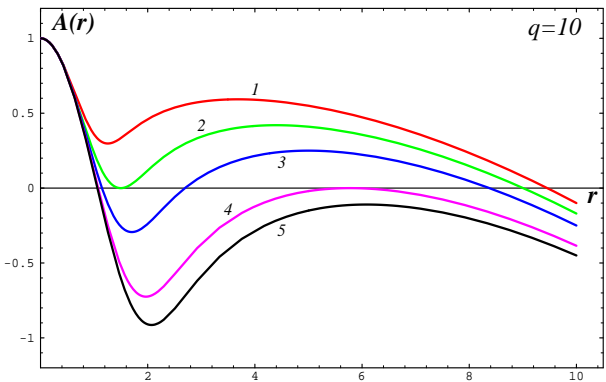,width=8.0cm,height=5.0cm}
\end{center}
\caption{Configurations described by $\Lambda\mN$.}
\label{fig.4}
\end{figure}

    The global structure of space-time contains in this case cosmological T
    regions which are asymptotically de Sitter as $r \to \infty$. The
    particular structure depends on the number and nature of horizons, which
    in turn depend on the values of the parameters in (\ref{rho}).

    According to \eqs (\ref{e11}) and (\ref{cons}), the transversal pressure
    $p_\bot$ can be expressed in terms of the function $A(r)$:
\beq
     8\pi G p_\bot = \Half A'' + \frac {A'}{r}.              \label{p-bot}
\eeq
    At an extremum of $A(r)$, $A'=0$, hence, if $p_\bot > 0$, this extremum
    is a minimum, and this minimum of $A$ is unique in the domain where
    $p_\bot> 0$ (otherwise there would be a maximum between two minima).
    Assuming that $p_\bot$ becomes negative only on distinguished
    length scales $l_i$ related to particular symmetry breakings with a de
    Sitter-like (false vacuum) behavior involved, we can fix the maximum
    number of horizons. One scale $l_1$ is related to the de Sitter core
    near the center. If there is no other symmetry breaking scale, we have
    an asymptotically flat configuration where $A(r)$ is positive at both
    small and large $r$ and has only one minimum, hence no more than two zeros
    (horizons) \cite{id02}. If there is a de Sitter asymptotic with small
    $\lambda$, this gives another scale $l_2$. Then there can be again only
    one domain $p_\bot >0$, but now $A$ has different signs at $r\to 0$ and
    $r\to \infty$, hence the single minimum of $A$ leads to at most 3
    horizons: an internal horizon $r_{-}$, a black hole horizon $r_{+}$, and
    a cosmological horizon $r_{++}$.

    Let us specify the possible types of cosmological models corresponding
    to globally regular configurations described by the cosmological term
    $\Lambda\mN$.

    All $\Lambda\mN$-dominated cosmologies belong to the \Lem\ class
    of anisotropic fluid models.  The cosmological expansion starts with a
    nonsingular non-simultaneous bang which we consider in detail in the
    next section.

    For a certain class of observers in T regions (specified in Sec. IIIC),
    $\Lambda\mN$-dominated cosmologies can also be identified
    as Kantowski-Sachs models with a regular R region.

    In cases 1 and 5 the global structure of space-time is the same as for
    the de Sitter geometry, but now the dynamics is governed by a variable
    cosmological term. The difference between the models of types
    1 and 5 is related to the particular dynamics. In case 1 the main
    dynamics occurs in the R region, and in case 5 in the T region.

    In cases 2, 3 and 4 in \fig 4 the global structure is more
    complex and is depicted in Figs.\,5--7 \cite{dg}. Particles moving in
    these geometries from the center to the asymptotically de Sitter T
    regions have to cross two horizons in cases 2 and 4, three horizons
    in case 3, and in cases 3 and 4 they have to cross an intermediate T
    region.

    In the case of three horizons the global structure is shown in Fig.\,5.
    (For a static observer in the R region between
    the black hole horizon $r_{+}$ and the cosmological horizon $r_{++}$,
    it corresponds to a nonsingular cosmological black hole \cite{us97}).
\begin{figure}
\vspace{-2.0mm}
\begin{center}
\epsfig{file=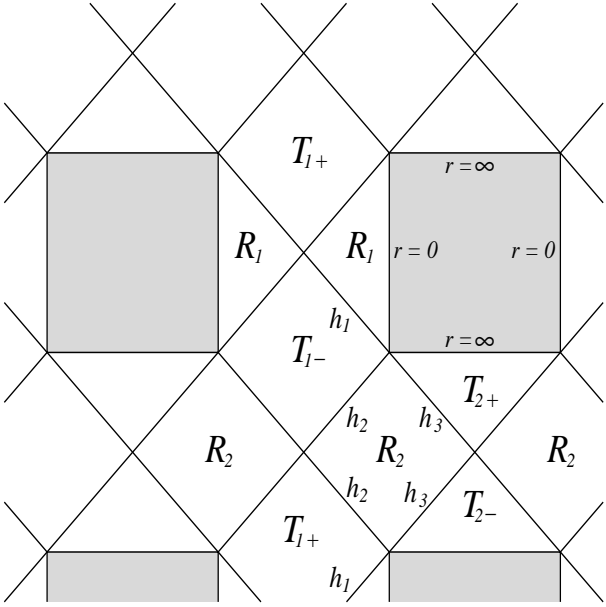,width=8.0cm,height=8.0cm}
\end{center}
\caption{Global structure of the $\Lambda\mN$ geometry
        with three horizons.}
\label{fig.5}
\end{figure}

    The global structure of space-time for the case when the black hole
    horizon $r_+$ coincides with the internal horizon $r_{-}$ is shown
    in \fig 6.
\begin{figure}
\vspace{-2.0mm}
\begin{center}
\epsfig{file=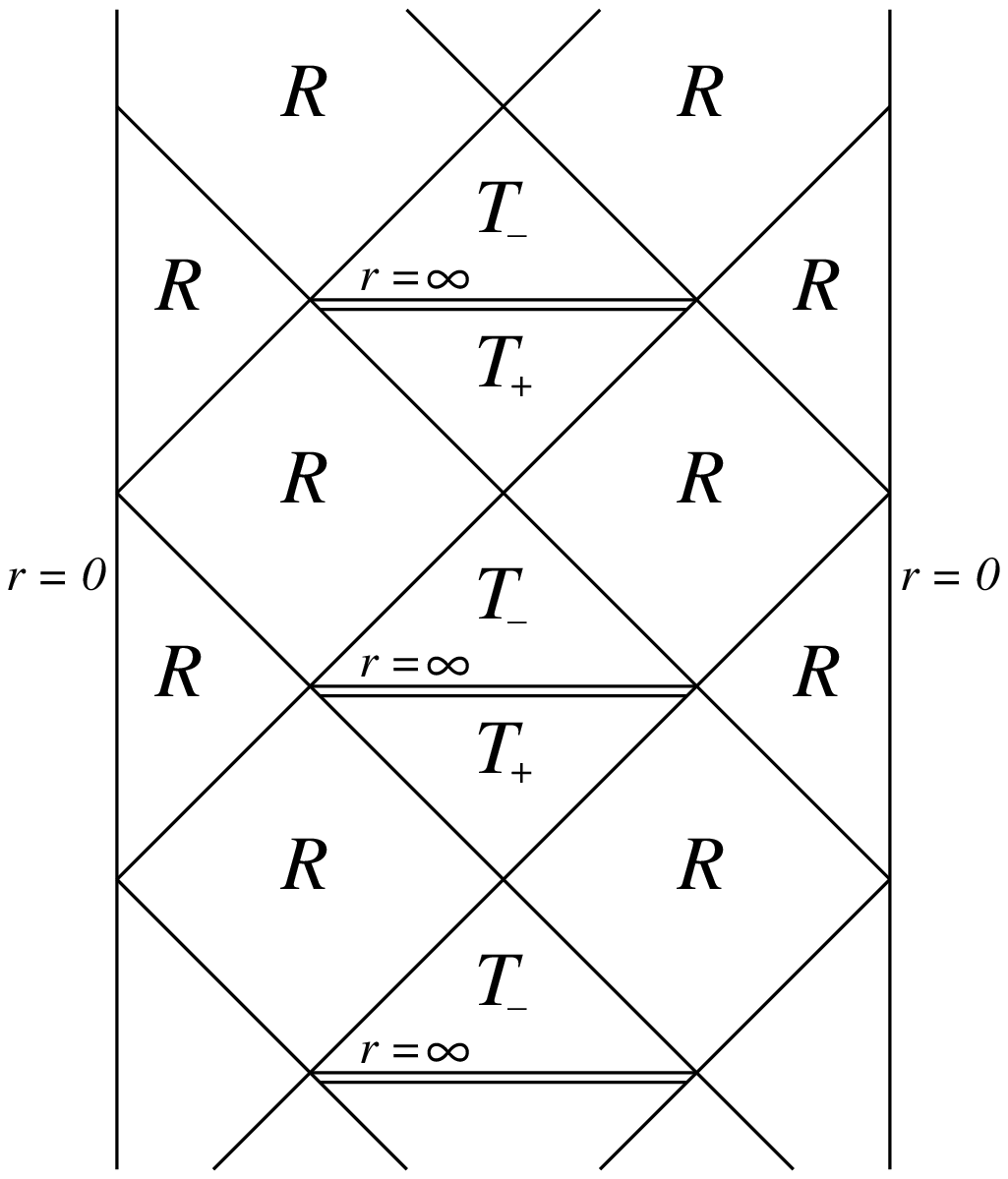,width=7.2cm,height=8.1cm}
\end{center}
\caption{Global structure of the $\Lambda\mN$ geometry
        with the double horizon $r_{-}=r_{+}$.}
\label{fig.6}
\end{figure}

    When the black hole horizon coincides with the
    cosmological horizon, the global structure is shown in Fig.\,7.
\begin{figure}
\vspace{-3.0mm}
\begin{center}
\epsfig{file=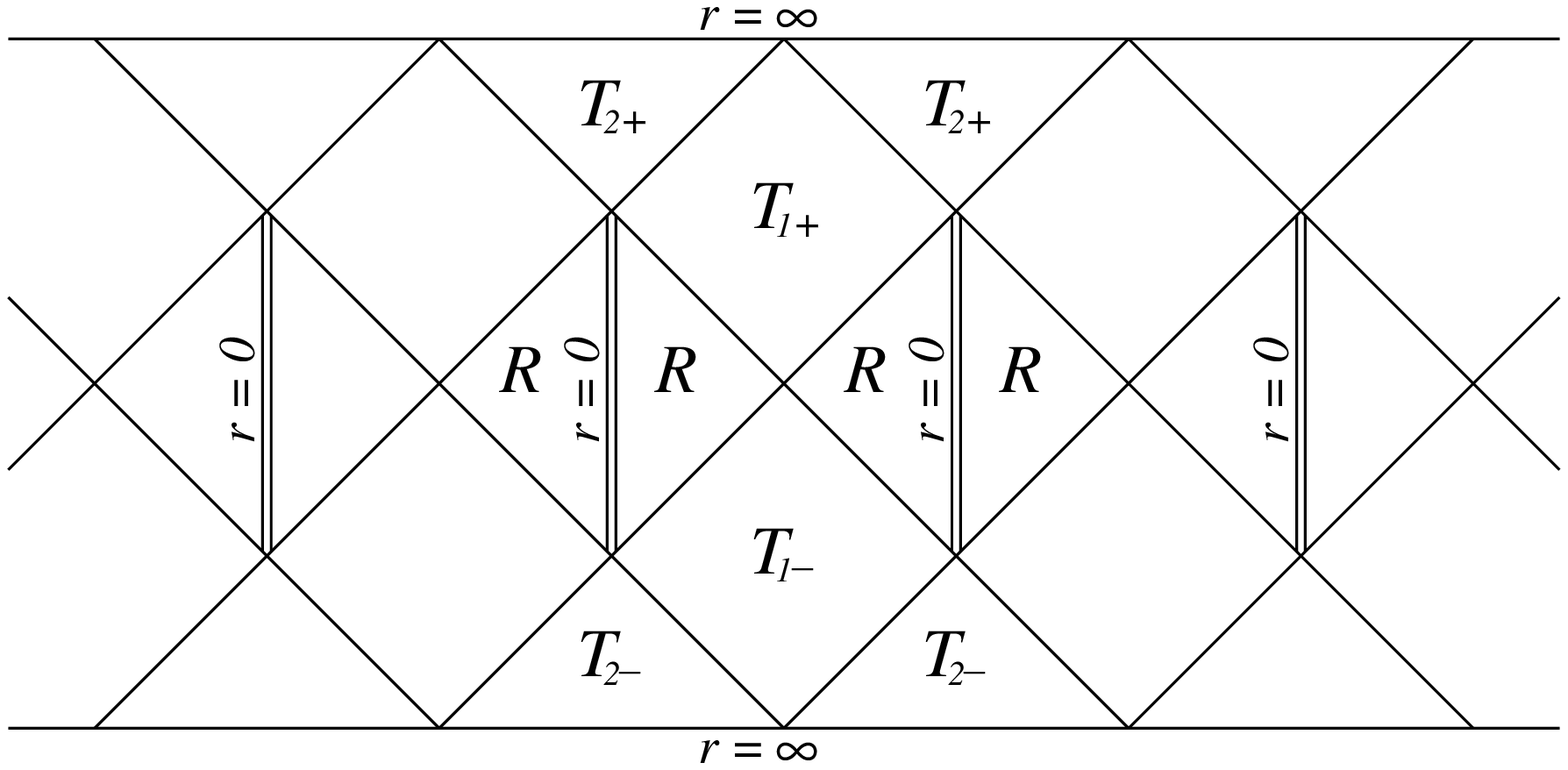,width=8.0cm,height=4.4cm}
\end{center}
\caption{Global structure of the $\Lambda\mN$ geometry with
    the double horizon $r_{+}=r_{++}$.}
\label{fig.7}
\end{figure}

    It is also possible that $A(r)$ has a triple horizon;
    this gives again the de Sitter space-time structure.

    In all these models, the cosmological evolution for a Kantowski-Sachs
    observer in the T region starts from a horizon, i.e., a null surface
    with vanishing volume of the spatial section.  We therefore call it
    a {\it null bang}. In the case of a double horizon a null bang
    occurs in the infinite past of an observer. In cases 1 and 5, as well as
    in the de Sitter case, the evolution starts from a simple horizon and
    at finite time from the point of view of a Kantowski-Sachs observer.

    The cosmological properties of these models are discussed in detail in
    the next sections.

\subsection {\Lem\ cosmologies with a nonsingular, nonsimultaneous bang}

     To present the $\Lambda\mN$ geometry as a \Lem-type \sph\ model,
     we connect an expanding reference frame with a certain congruence of
     radial geodesics and consider this frame as comoving to our
     cosmological model. The procedure is similar to what Lema\^\i tre
     has done for the \Sch\ metric \cite{ll}. The algebraic structure
     (\ref{Tstru}) of $\Lambda\mN$ implies that a reference frame connected
     with any radial motion will automatically be comoving to our vacuum
     matter.

     A transition to the geodesic coordinates $(R,\tau)$, where $\tau$ is
     the proper time along a geodesic and the radial coordinate $R$ is the
     congruence parameter, different for different geodesics, can be
     described in a general form. A radial timelike geodesic in the metric
     (\ref{g_sph}) satisfies the equations
\beq
     \biggl(\frac{dr}{d\tau}\biggr)^2 = E^2 - A(r),     \qquad
     \frac{dt}{d\tau} = \frac {E}{A(r)},                     \label{geo'}
\eeq
     where the constant $E$ is connected with the initial velocity of a
     particle moving along this particular geodesic at a given value of $r$.
     Relating the reference frame to geodesics with a certain fixed value of
     $E$, we can perform a transition from the coordinates $(r,t)$ to the
     coordinates ($R,\tau$) with the following transformation:
\bear                                                        \label{trans}
      \d_\tau r \eql \pm \sqrt{E^2 -A}, \qquad
      \d_R r = \sqrt{E^2 -A},
\\
      \d_\tau t \eql \frac{E}{A}, \cm\
      \cm \d_R t = \pm\frac{E^2-A}{AE},
\ear
     where the plus and minus signs refer to growing and falling $r(\tau)$
     (expanding and contracting models), respectively. The resulting metric
     has the form
\beq                                                       \label{ds_geo}
     ds^2 = d\tau^2 - \frac{E^2 - A(r)}{E^2} dR^2
                        -r^2 (R,\tau)d\Omega_1^2,
\eeq
     where $r(R,\tau)$ should be determined from \eqs (\ref{trans}).
     For expanding models, since $\d_\tau r = \d_R r$, we see that $r$ is a
     function of $R+\tau$. (In this construction we have used the freedom of
     re-parametrization $R \mapsto \tilde R (R)$.)

     Such a model for de Sitter-Schwarzschild geometry has been
     considered in Ref.\cite{us2001} for radial geodesics with $E=1$ in
     the present notations. According to (\ref{geo'}), the condition $E=1$
     means that the velocity of particles realizing our reference frame
     is zero at $r=0$. As a result, the center $r=0$, where the expansion
     starts, corresponds to the surface $R+\tau=-\infty$ in the coordinates
     $(R,\tau)$. Different points of the bang surface $r(R,\tau)=0$
     \cite{silk} (which is time-like in $\Lambda\mN$ geometry) start at
     different instants of synchronous time $\tau$. In the limit $r\to 0$
     the metric transforms to the de Sitter metric
\beq
     ds^2 = d\tau^2 - \e^{2H\tau} (dx^2 + x^2 d\Omega_1^2)    \label{RW_0}
\eeq
     with $H =\sqrt{\Lambda/3}$ and $x = H^{-1} \e^{HR}$ and describes
     a nonsingular nonsimultaneous de Sitter bang \cite{us2001},
     followed by a Kasner-type anisotropic stage, with contraction
     in the radial direction and expansion in the tangential direction, at
     which most of the Universe mass is produced; the metric
     at the intermediate stage has the form \cite{us2001}
\beq
     ds^2 = d\tau^2- (\tau {+} R)^{-2/3} F(R) dR^2 -
       \label{kasner}
                         B(\tau {+} R)^{4/3} d\Omega_1^2
\eeq
     where $F(R)$ is a smooth regular function and $B$ is a constant
     related to the model parameters.

     Similar models can be constructed in the general case with one,
     two or three Killing horizons. In all $\Lambda_{\mu}^{\nu}$ geometries
     the regular world line $r=0$ is timelike (see Figs.\,3, 5),
     therefore different particles that realize our reference frame leave
     or cross this world line at different instants of synchronous time
     $\tau$. The start of the expansion is therefore the same in the
     general case, i.e., a nonsingular nonsimultaneous de Sitter bang at
     $r=0$, occuring at $R + \tau\to -\infty$ for $E=1$ and at finite $R +
     \tau$ if $E > 1$, when the initial velocity of the geodesic reference
     frame is nonzero. This inflationary stage is followed by a highly
     anisotropic stage of evolution. Expansion in the tangential direction
     ($\d_\tau r >0$) is accompanied by contraction in the radial direction
     ($\d_\tau |g_{RR}| <0$) as long as $dA/dr > 0$, as is easily seen
     from (\ref{ds_geo}). At large values of $R+\tau$, when the
     metric achieves another de Sitter asymptotic corresponding to the
     cosmological constant $\lambda$, the model becomes isotropic and is
     again described by the metric (\ref{RW_0}), but now with the Hubble
     parameter $H = \sqrt{\lambda/3}$.

\subsection{Kantowski-Sachs cosmologies with a null bang}

    The asymptotically de Sitter regions of the above two-lambda
    configurations represent, as any \sph\ T regions, KS cosmological models.
    The metric has the form (\ref{dsT}) with the temporal coordinate $u=r$
    and $b^2(u)=|A(r)|$:
\beq
    ds^2 = \frac{1}{b^2(r)}dr^2 - b^2(r)dx^2 - r^2 d\Omega_1^2  \label{KS}
\eeq
    The KS evolution at late times corresponds to the de Sitter large $r$
    asymptotic (\ref{dS+}). In all cases the KS evolution starts with a null
    bang from a horizon. This initial state with a finite value of $r$ is,
    from the viewpoint of comoving observers in the model (\ref{KS}), a
    highly anisotropic, purely coordinate singularity where the spatial
    sections, having the topology of a 3-dimensional cylinder, are squeesed
    along the ``longitudinal'' direction $x$ due to vanishing $b(r)$. The
    4-geometry is, however, globally regular.

    In cases 1, 2, 3 and 5 according to \fig 4, this initial state occurs
    at finite proper time $\tau = \int {dr/b(r)}$. Thus our KS models behave
    qualitatively as the de Sitter model (see \sect 2.2), but with
    quite different prehistories. In case 4, with a double horizon
    connecting two T regions, the null bang occurs in the remote past,
    $\tau\to -\infty$. The same is true for the case of a triple horizon.

    Let us show that observers who follow other geodesics  cross this
    past horizon at finite instants of their proper times. Indeed, the
    geodesic equations in any space-time with the metric (\ref{ds}) have the
    following integral:
\beq
      (dr/d\tau)^2 + k_1 A + L^2 A/r^2 = E^2               \label{geo}
\eeq
    where $E$ and $L$ are the constants of motion associated with the
    particle energy and angular momentum, while $k_1=1$ and $k_1=0$ for
    timelike and null geodesics, respectively; the affine parameter $\tau$
    has the meaning of proper time along the geodesic in case $k_1=1$. At a
    horizon $r=h$, $A(r)$ vanishes, and in case $E > 0$ one has $dr/d\tau\ne
    0$ for all geodesics, whence it follows that $|\tau| < \infty$,
    irrespective of the order of the horizon. The time lines in the T region
    which are trajectories of the KS observers, evidently correspond to
    $k_1=1$, $L=0$, $E=0$, so that for these lines we return to the relation
    (\ref{tau1}), leading to infinite $\tau$ at horizons of order two and
    higher.

    Therefore even ``slow'' observers whose world lines coincide with
    these time lines receive information from their infinitely remote past,
    coming with particles and photons which have crossed the horizon.

    The reasoning related to \eq (\ref{geo}) is quite general, and
    in other KS cosmologies, starting with a null bang, the prehistory is
    also observable.

    One can take as a simple example the \Sch-de Sitter metric
    (\ref{g_sph}), (\ref{SdS}) with $K=1$, $m > 0$ and $\lambda > 0$.
    The metric function $A(r) = 1 - 2Gm/r - (\lambda/3) r^2$ is plotted in
    Fig.\,8.

\begin{figure}
\vspace{-8.0mm}
\begin{center}
\epsfig{file=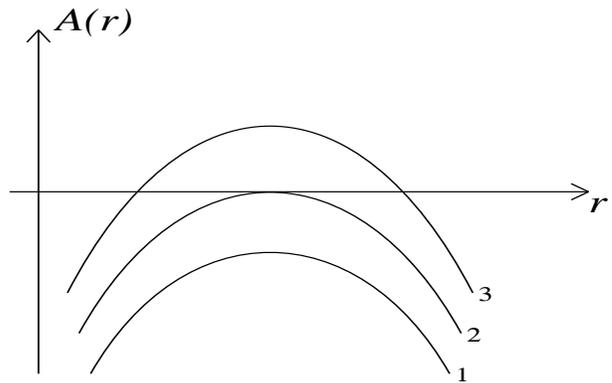,width=8.0cm,height=5.0cm}
\end{center}
\caption{The function $A(r)$ for the Schwarzschild-De Sitter geometry}
\label{fig.8}
\end{figure}

    Three cases should be distinguished:

    (i) $3Gm > 1/\sqrt{\lambda}$, no horizon (curve 1). It is a global KS
    model with a highly anisotropic singularity $r=0$ in the past and a de
    Sitter future asymptotic.

    (ii) $3Gm = 1/\sqrt{\lambda}$, one double horizon $h = 3Gm$,
    separating two T regions (curve 2). The region $r>h$ is similar
    to the large $r$ region in case 4 of the $\Lambda\mN$ geometry: a null
    bang in the infinitely remote past.

    (iii) $3Gm < 1/\sqrt{\lambda}$, two horizons $h_1$ and $h_2$ (curve 3).
    The region $r <h_1$ is quite similar to the \Sch\ \bh\ interior, while
    the region $r > h_2$ has the same qualitative features as
    the large $r$ region in cases 1, 2, 3 and 5 of the $\Lambda\mN$ geometry
    (\fig 4), and the null bang takes place at finite proper time.

    As before, observers in the KS regions receive signals from
    the pre-historical inner regions. Thus in \Sch-de Sitter
    space-time the past anisotropic singularity is visible to all observers.

    One more case of interest for comparison (though with a different
    structure of the SET) is that of a scalar field minimally coupled to
    gravity, with the Lagrangian
\beq                            \label{Lagr}
     L_\varphi = g\MN\varphi_{,\mu}\varphi_{,\nu} -2V(\varphi),
\eeq
    where $V(\varphi)$ is a potential. This field with various potentials
    is applied to various cosmological and local problems. If $\varphi$ is
    constant, the system reduces to vacuum in \GR\ with the cosmological
    constant given by the corresponding value of $V$=const. As shown in
    \Ref{vac1}, in the case of \ssph\ configurations, a variable scalar
    field adds nothing to the list of possible global space-time structures
    with constant $\varphi$, whatever is the choice of the function
    $V(\varphi)$ (not necessarily positive-definite). In particular,
    there can be no more than two horizons; the only possible double horizon
    separates two T regions. If there is a \bh\ horizon (i.e., a horizon
    $r=h$ separating a T region at $r<h$ from an R region at $r>h$), there
    is always a singularity at $r=0$. The system with any small $r$
    behaviour has a de Sitter asymptotic at infinity if $\varphi\to
    \varphi_0$ such that $V(\varphi_0) >0$ at large $r$.

    A common feature with the $\Lambda\mN$ geometry is that scalar field
    configurations which have a regular center are always asymptotically de
    Sitter at $r=0$, but in the scalar case they can have either no horizon,
    or one cosmological horizon.

    Therefore the above discussion of regular $\Lambda\mN$ geometries
    (cases 1 and 5 according to \fig 4) is applicable to regular scalar
    field configurations with a cosmological horizon. The discussion of
    the \Sch-de Sitter geometry is applicable to the class of systems with
    scalar fields, singular at $r=0$.

    A qualitative picture similar to cases 1 and 5 according to \fig 4 is
    also observed if a single scalar field is replaced with a scalar triplet
    forming a global monopole configuration \cite{vilshel}: in a certain
    range of the gravitational field strength parameter, the monopole has a
    de Sitter core, separated by a cosmological horizon from an outer KS
    region with a large $r$ asymptotic, characterized by a zero or small
    cosmological constant \cite{glomon}.

\section {Summary and discussion}

    Let us briefly summarize our results.

    We have presented regular $\Lambda\mN$ dominated perfect fluid models in
    which the evolution starts from the de Sitter vacuum $\Lambda \delta\mN$
    with $\Lambda$ on the scale of symmetry breaking (which can be a Grand
    Unification or SUSY scale) and ultimately approaches another de Sitter
    vacuum $\lambda \delta_{\mu}^{\nu}$ with $\lambda < \Lambda$.

    These models belong to the \Lem\ class of anisotropic fluid models.
    The cosmological expansion starts from a nonsingular non-simultaneous
    de Sitter bang, followed by a Kasner type anisotropic stage and
    approaches isotropic FRW expansion at late times.

    Exact solutions giving rise to $\Lambda\mN$ dominated models
    are presented for three symmetries (spherical, planar and
    pseudospherical) of spatial 2-surfaces. All  planar and pseudospherical
    regular models are anisotropic T models (without R regions) with
    isotropization at late times.  The planar case is classified as the
    Bianchi I type and the pseudospherical case can be called the hyperbolic
    Kantowski-Sachs class.

    In the spherically symmetric case the models contain regular R regions
    and can be identified as Kantowski-Sachs models with regular R regions.

    A remarkable feature of these models is that, for a Kantowski-Sachs
    observer, the evolution starts with a ``null bang", from a null surface
    which seems singular to comoving observers but which is perfectly
    regular in the 4-dimensional $\Lambda\mN$ geometry. Information about
    the pre-null bang history is available to Kantowski-Sachs observers.

    The KS and \Lem\ type cosmologies presented here are built on the
    basis of the same space-times but differ in the choice of reference
    frames. This is an example of the evident circumstance that the
    observational properties of space-time in \GR\ depend on the observers'
    motion, in other words, on the choice of a reference frame.
    Another, well-known example is the de Sitter metric which, being written
    in different reference frames, gives rise to all three types of FRW
    isotropic models. Two anisotropic representations of the de Sitter
    geometry are given here in \eqs (\ref{dS+}) and (\ref{dS-}).

    We are currently working on detailed cosmological models consistent
    with all observational constraints. Our preliminary results testify that
    $\Lambda\mN$ cosmologies are able to provide a smooth decay of vacuum
    density by more than hundred orders of magnitude during the cosmological
    evolution.

\subsection*{Acknowledgment}

     This work was supported by the Polish Committee for Scientific Research
     through grant No. 5P03D.007.20 and a grant for UWM.

     KB thanks the colleagues from the University of Warmia and Mazury,
     where part of the work was done, for kind hospitality and acknowledges
     partial support from the Ministry of Industry, Science and
     Technologies of Russia.


\begin{thebibliography}{99}
\bibitem{dg}
    These diagrams have been first built for the Reissner-Nordstr\"om-de
    Sitter geometry \cite{kb79}, but now, just as in the asymptotically
        flat case, we have a regular de Sitter core instead of a singularity
        at $r=0$.
\end{thebibliography}
\end{document}